\documentclass[sensors,letter,submit,moreauthors,pdftex]{definitions/mdpi} 

\firstpage{1} 
\pubvolume{xx}
\issuenum{1}
\articlenumber{5}
\pubyear{2020}
\copyrightyear{2020}
\history{Received: date; Accepted: date; Published: date}

\pdfoutput=1
\preto{\abstractkeywords}{\nolinenumbers}

\usepackage{amssymb}
\usepackage{amsthm}
\usepackage{bm}

\usepackage{setspace}
\usepackage{booktabs}
\usepackage[flushleft]{threeparttable}
\usepackage{array}
\usepackage{ragged2e}
\newcolumntype{L}[1]{>{\RaggedRight\hspace{0pt}}p{#1}}
\usepackage[ruled,vlined]{algorithm2e}

\newtheorem{remarkenv}{Remark}

\newcommand{\bmath}[1]{{\mathbf{#1}}}
\newcommand{\setC}{{\mathbb{C}}}
\newcommand{\setR}{{\mathbb{R}}}

\newcommand{\eqdef}{\triangleq}
\newcommand{\herm}{\text{H}}
\newcommand{\trasp}{\text{T}}
\newcommand{\Es}{{\mathbb{E}}}          % expectation
\newcommand{\rank}{{\text{rank}}}
\newcommand{\diag}{{\text{diag}}}
\newcommand{\trace}{{\text{tr}}}

\newcommand{\I}{\bmath{I}}

\newcommand{\Nc}{N_{\text{C}}}
\newcommand{\Nr}{N_{\text{R}}}

\newcommand{\Ns}[1]{N_{\text{S},#1}}
\newcommand{\Ntone}{N_{\text{T}}}
\newcommand{\Nt}[1]{N_{\text{T},#1}}

\newcommand{\z}{\bmath{z}}
\renewcommand{\S}{\bmath{S}}
\newcommand{\A}{\bmath{A}}

\newcommand{\B}{\bmath{B}}
\newcommand{\E}{\bmath{E}}

\renewcommand{\C}{\bmath{C}}
\newcommand{\D}{\bmath{D}}
\newcommand{\Ua}{\bmath{U}_{\text{a}}}
\newcommand{\Lambdaa}{\bm{\Lambda}_{\text{a}}}

\newcommand{\Dmmsei}{\bmath{D}_{i,\text{mmse}}}
\newcommand{\Q}{\bmath{Q}}
\renewcommand{\U}{\bmath{U}}
\newcommand{\V}{\bmath{V}}
\newcommand{\e}{\bmath{e}}

\renewcommand{\P}{\bmath{P}}
\newcommand{\w}{\bmath{w}}

\newcommand{\n}{\bmath{n}}
\renewcommand{\r}{\bmath{r}}

\newcommand{\s}{\bmath{s}}

\newcommand{\x}{\bmath{x}}
\newcommand{\y}{\bmath{y}}
\newcommand{\F}{\bmath{F}}
\renewcommand{\H}{\bmath{H}}
\renewcommand{\G}{\bmath{G}}
\renewcommand{\v}{\bmath{v}}

\newcommand{\Rvvi}{\bmath{K}_{\v_i\v_i}}
\newcommand{\Rvvinot}{\bmath{K}_{\v_{\inot}\v_{\inot}}}
\newcommand{\Ryy}{\bmath{K}_{\y\y}}

\newcommand{\Reei}{\bmath{K}_{\e_i\e_i}}
\newcommand{\Pk}{\mathcal{P}_{k}}

\newcommand{\Ptx}[1]{\mathcal{P}_{\text{T},#1}}
\newcommand{\Prx}[1]{\mathcal{P}_{\text{R},#1}}
\newcommand{\Prxtilde}[1]{\widetilde{\mathcal{P}}_{\text{R},#1}}

\newcommand{\inot}{\underline{i}}

\def\bdm#1\edm{\begin{displaymath}#1\end{displaymath}}
\def\be#1\ee{\begin{equation}#1\end{equation}}
\def\barr#1\earr{\begin{align}#1\end{align}}

\Title{Separable MSE-based design of two-way multiple-relay cooperative MIMO 5G networks}

\Author{Donatella Darsena$^{1,4}$\orcidA{},
Giacinto Gelli$^{2,4}$,
Ivan Iudice$^{3}$\orcidC{},
and Francesco Verde$^{2,4}$\orcidD{}}

\AuthorNames{Donatella Darsena, Giacinto Gelli, Ivan Iudice,
and Francesco Verde}

\address{%
$^{1}$ \quad Department of Engineering, Parthenope University, Naples I-80143, Italy \\
$^{2}$ \quad Department of Electrical Engineering and
Information Technology, University of Naples Federico II, Italy \\
$^{3}$ \quad Italian Aerospace Research Centre (CIRA),
Capua I-81043, Italy \\
$^{4}$ \quad National Inter-University Consortium for Telecommunications (CNIT)
}

\corres{Correspondence: f.verde@unina.it (F.V.)}

\abstract{
While the combination of multi-antenna and relaying techniques 
has been extensively studied for  Long Term Evolution
Advanced (LTE-A) and Internet of Things (IoT) applications, 
it is expected to still play an important role in 5th Generation (5G)
networks. However, the expected benefits of these technologies  
cannot be achieved without a proper system design.
In this paper,   
we consider the problem of jointly optimizing
terminal precoders/decoders and
relay forwarding matrices on the basis of the sum mean square error (MSE) criterion in multiple-input multiple-output (MIMO) two-way relay systems, where two multi-antenna nodes mutually exchange information via multi-antenna amplify-and-forward relays.
This problem is nonconvex and a local optimal solution is typically found by using iterative algorithms based on alternating optimization.
We show how the constrained minimization of the sum-MSE can be relaxed to
obtain two separated subproblems which, under mild conditions, admit a
closed-form solution.
Compared to iterative approaches, the proposed design is more suited to 
be integrated in 5G networks, since it is 
computationally more convenient and its 
performance exhibits a better scaling in the number of relays. 
}


\begin{document}

\section{Introduction}
\label{sec_intro}

Cooperative multiple-input  multiple-output (MIMO)
communication techniques, wherein data exchange
between MIMO terminal nodes
is assisted by one or multiple MIMO relays,
have been studied for 
Long Term Evolution Advanced (LTE-A) 
cellular systems \cite{Yang2009,Loa2010,Bhat2014},  
since they assure significant
performance gains in terms of
coverage, reliability, and capacity. 
Relay technology has been also considered for Internet of Things (IoT) applications, 
by allowing in particular the support of the massive access for fog and social networking services
\cite{Omo2019,Chen2020, Ji2020}.
One of the main changes when going from LTE-A to 5th
generation (5G) systems is the spectrum use at radically higher frequencies 
in the millimeter-wave (mmWave) range \cite{Parkwall2017}.
However, mmWave signals are  highly  susceptible not only to 
blockages  from  large-size  structures,  e.g.,  buildings, but they are also   
severely  attenuated  by the presence of small-size objects, e.g., human bodies and foliage \cite{Rap2019}.
In this regard, cooperative MIMO technology additionally represents 
a possible  approach  for  circumventing  the  unreliability  of 
mmWave channels \cite{Wu2018} in 5G networks. 

In addition, 5G systems have stringent requirements
in terms of spectral efficiency. 
Many relaying protocols operate in
\emph{half-duplex} mode \cite{Cui2004,Jay2006,Beh2008,Dar2019},
where two time slots are required
to perform a single transmission,
due to the inability of the relays to receive and
transmit at the same time.
To overcome the inherent
halving of spectral efficiency,
a possible remedy for 5G applications
is to adopt \emph{two-way} 
relaying \cite{Katti2007} (see Fig.~\ref{fig:fig_1}), which works as follows:
(i) in the first slot, the two terminal nodes
simultaneously transmit
their precoded signals to the relays;
(ii) in the second slot, the relays precode and
forward the received signals to the terminals.
Since each terminal knows its own transmitted signal,
the effects of self-interference can be subtracted from
the received signal at the terminals,
and the data of interest can be decoded.
On the other side of the coin, with respect to the one-way relaying setting,
the optimization of two-way cooperative networks is complicated by
fact that terminal precoders/decoders and relay forwarding matrices 
are coupled  among themselves. 

Design and performance analysis
of two-way cooperative MIMO networks encompassing multiple
\emph{amplify-and-forward} (AF)
or \emph{non-regenerative}
relays has been considered in 
\cite{Lee2010,Ron2011,Hu2013,Hu2016a,Hu2018a}.
Compared with the single-relay case \cite{Ron2012},
the multiple-relay scenario
generally leads to
more challenging \emph{nonconvex} constrained optimization problems,
which are usually solved by burdensome iterative procedures.
In \cite{Lee2010}, by adopting a
weighted sum-mean-square-error (MSE)
or a sum-rate cost function,
iterative gradient descent optimization
algorithms are proposed, with transmit-power
constraints imposed  at both  the terminals
and the relays.
A similar scenario is
considered in \cite{Ron2011} and \cite{Hu2013}.
In \cite{Ron2011}, 
the original constrained minimum sum-MSE
nonconvex  optimization problem is iteratively solved. 
Specifically, the algorithm of \cite{Ron2011} starts by 
randomly choosing  the terminal precoders and the
relay forwarding matrices  satisfying the 
transmission power constraints at the source terminals 
and the relay nodes. In each iteration, 
the terminal precoders, the relay forwarding matrices, and
the decoders are alternatingly updated in \cite{Ron2011} through solving 
convex subproblems:  first, with given precoders and
relay forwarding matrices, the optimal decoders 
are obtained in closed-form by solving an unconstrained convex problem; 
second, with fixed precoders and decoders, the relaying matrix of all
the relays are updated in closed-form one-by-one by freezing 
the relaying matrices of the other relays; 
finally, given the decoders and relaying matrices, the precoders 
are updated by solving a convex quadratically constrained quadratic programming 
problem.
A different iterative optimization procedure is proposed in \cite{Hu2013},
based on the matrix conjugate gradient algorithm,
which is shown to converge
faster than conventional
gradient descent methods.
Finally, some recent papers
\cite{Hu2016a,Hu2018a}
propose architectures for two-way relaying
based on relay/antenna selection strategies.

In this paper, 
we propose an optimization algorithm for two-way
AF MIMO relaying 5G networks,
where terminal precoders/decoders and
relay forwarding matrices are jointly
derived under power constraints
on the transmitted/received power
at the terminals.
Rather than attempting to solve
it iteratively, we
derive a relaxed version of the
original minimum sum-MSE nonconvex optimization,
which allows one to decompose it in two
separate problems that  admit
a closed-form, albeit suboptimal,
solution.
We show by Monte Carlo trials
that our closed-form approach performs comparably or better than representative
iterative approaches proposed in the literature for the same scenario
with a reduced computational complexity,
especially for increasing values of the number of relays.

%%%%%%%%%%%%%%%%%%%%%%%%%%%%%%%%%%%%%%%%%%
\section{Network model and basic assumptions}
\label{sec:model}

\begin{figure}[t]
\begin{center}
\includegraphics[width=\textwidth]{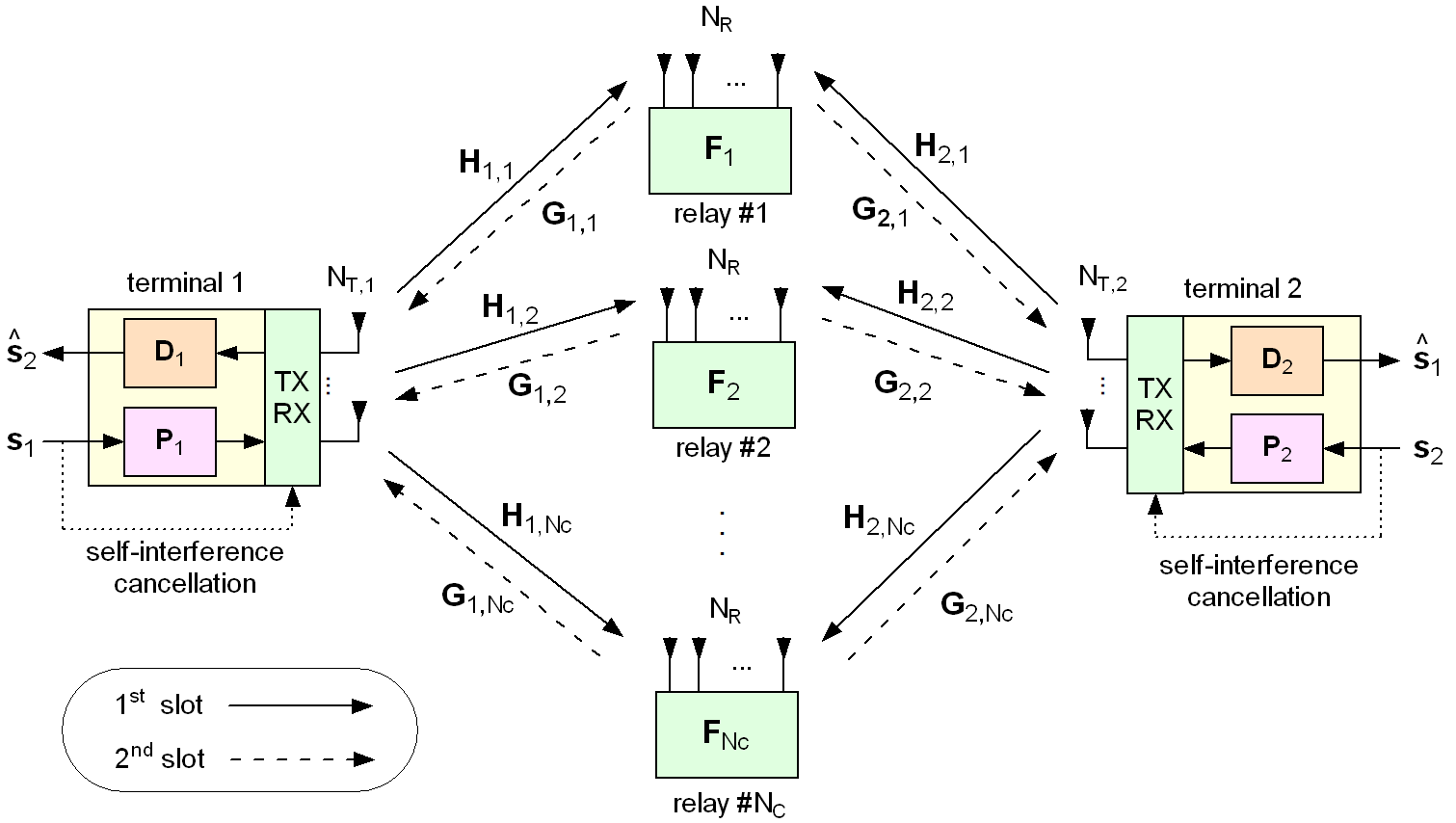}
\caption{Model of the considered two-way relaying MIMO 5G network.}
\label{fig:fig_1}
\end{center}
\end{figure}

We consider the two-way MIMO 5G network configuration of
Fig.~\ref{fig:fig_1}, where bidirectional
communication between two terminals, equipped with
$\Nt1$ and $\Nt2$ antennas, respectively,
is assisted
by $\Nc$ half-duplex relays, each equipped
with $\Nr$ antennas.
We assume that there is no direct link between the two terminals,
due to high path loss values or obstructions.
Even though our approach can be generalized, for simplicity,
the considered physical layer is that of a single-carrier cooperative system
where all the channel links are quasi static and experience flat fading.

Let $\s_1 \in \setC^{\Ns1}$ and $\s_2\in \setC^{\Ns2}$
denote the symbol vectors
to be transmitted by terminal 1 and 2, respectively.
In the first time slot, each terminal
precodes its symbols with matrix
$\P_i \in \setC^{\Nt{i} \times \Ns{i}}$,
for $i \in \{1,2\}$,
before transmitting it to the relays,
which thus receive
$\y_k = \sum_{i=1}^{2} \H_{i,k} \P_i \, \s_i + \w_k$, 
for $k \in \{1, 2, \ldots, \Nc\}$,
where $\H_{i,k} \in \setC^{\Nr \times \Nt{i}}$
is the \emph{first-hop} channel
matrix
(from terminal $i$ to relay $k$),
and $\w_k \in \setC^{\Nr}$ models additive  noise
at $k$th relay.
By defining
$\y \eqdef [\y_1^T, \y_2^T, \ldots, \y_{\Nc}^T]^T \in \setC^{\Nc\Nr}$,
the overall signal received by the relays can be compactly
written  as
\be
\y = \sum_{i=1}^{2} \H_{i} \P_i \, \s_i + \w
\label{eq:rel-2}
\ee
where $\H_i \eqdef [\H_{i,1}^T, \H_{i,2}^T \ldots,   \H_{i,\Nc}^T]^T \in \setC^{\Nc\Nr \times \Nt{i}}$ gathers all first-hop channels and the vector 
$\w \eqdef [\w_1^\trasp, \w_2^\trasp, \ldots, \w_{\Nc}^\trasp]^\trasp \in \setC^{\Nc \Nr}$
gathers all the noise samples. 

In the second time slot, the $k$th relay forwards
its received signal $\y_k \in \setC^{\Nr}$,
by using the relaying matrix
$\F_k \in \setC^{\Nr \times \Nr}$, thus
transmitting 
$\z_k = \F_k \, \y_k$.
The received signal at each terminal can be written,
for $i \in \{ 1,2 \}$, as
\be
\r_i = \sum_{k=1}^{\Nc} \G_{i,k} \F_k \, \y_k + \n_i = \G_{i} \, \F \, \y + \n_i
\label{eq:rec-1}
\ee
where $\G_{i,k} \in \setC^{\Nt{i} \times \Nr}$ is the
\emph{second-hop} channel matrix (from relay $k$ to terminal $i$),
and the vector $\n_i \in \setC^{\Nt{i}}$ is additive noise at terminal $i$.
Additionally, we have defined  in \eqref{eq:rec-1} the extended matrices 
$\G_i \eqdef [\G_{i,1}, \G_{i,2} \ldots,   \G_{i,\Nc}]\in \setC^{\Nt{i}\times\Nc \Nr}$
and
$\F \eqdef \diag( \F_1, \F_2, \ldots, \F_{\Nc}) \in \setC^{\Nc\Nr \times \Nc\Nr}$.
Moreover, by taking into account \eqref{eq:rel-2}, the vector $\r_i$
can also be directly written in terms of $\s_1$ and $\s_2$ as
\be
\r_i = \sum_{j = 1}^{2} \C_{i,j} \, \s_j  + \v_{i}
\ee
where
$\C_{i,j} \eqdef \G_{i} \, \F \, \H_{j} \, \P_{j} \in \setC^{\Nt{i} \times \Ns{j}}$ is the \emph{dual-hop} matrix from terminal $j$ to $i$, for $i,j \in \{1,2\}$, and  vector
$\v_i \eqdef \G_i \,\F \, \w + \n_i \in \setC^{\Nt{i}}$ is the
overall noise.

We assume customarily \cite{Katti2007,Hu2016a} that each terminal can
estimate and subtract the self-interference
deriving from its own symbols. To do this, terminal $i$ 
has to first acquire the matrix $\mathbf{C}_{i,i}$, which can be obtained 
by resorting to standard training-based identification methods. 
Specifically, each
data transmission can be preceded by a training period, wherein
the two terminals transmit orthogonal pilot sequences
to the relays.
In this case, by redefining
$\r_i$ with a slight abuse of notation as $\r_i - \C_{i,i} \, \s_i$, for $i \in \{1,2\}$,
we write explicitly
\be
\r_i  = \C_{i,\inot} \, \s_{\inot}  + \v_i = \G_{i} \F \H_{\inot} \, \P_{\inot} \, \s_{\inot} + \v_i
\label{eq:basic-model}
\ee
where
$\inot = 2$  when  $i=1$, whereas
$\inot = 1$ when $i=2$.

At terminal $i$, vector $\r_i$
is subject to linear equalization
through matrix $\D_i \in \setC^{\Ns{\inot} \times \Nt{i}}$,
thus yielding a soft estimate
$\hat{\s}_{\inot} \eqdef \D_{i} \, \r_i$ of
the symbols $\s_{\inot}$
transmitted by terminal $\inot$,
whose entries are then
subject to minimum-distance hard decision.

In the sequel, we consider the
common assumptions:
({a}1)
$\s_1$ and $\s_2$ are mutually independent
zero-mean circularly symmetric
complex (ZMCSC) random vectors, with
$\Es[\s_i\, \s_i^{\herm}] = \I_{\Ns{i}}$, for $i \in \{1,2\}$;
({a}2)
the entries of $\H_i$ and $\G_i$ are independent
identically distributed
ZMCSC Gaussian unit-variance random variables, for $i \in \{1,2\}$;
({a}3)
the noise vectors $\w$, $\n_1$ and $\n_2$ are mutually independent
ZMCSC Gaussian random vectors, statistically independent of
$\{\s_i,\H_i,\G_i\}_{i=1}^{2}$, with
$\Es[\w \w^{\herm}] = \sigma^2_w \, \I_{\Nc\Nr}$ and
$\Es[\n_i \n_i^{\herm}] = \sigma^2_{n,i} \, \I_{\Nt{i}}$,
for $i \in \{1,2\}$.

Full channel-state information (CSI) is
assumed to be available
at both the terminals and the relays.
Particularly, we assume that:
(i) $\{\H_i\}_{i=1}^{2}$ are known
at the terminals and at the relays;
(ii) the $k$th second-hop channel matrices
$\G_{1,k}$ and $\G_{2,k}$
are known only to the $k$th relay,
for $k \in \{1,2,\ldots, \Nc\}$;
(iii) the
dual-hop channel matrix $\{\C_{i,\inot}\}$ and
the covariance matrix%
\footnote{Hereinafter all the ensemble averages are evaluated
for fixed values of the first- and second-hop
channel matrices.}
\be
\Rvvi \eqdef \Es[\v_i \, \v_i^\herm] = \sigma_w^2 \, \G_i \, \F \, \F^\herm \, \G_i^\herm 
+  \sigma^2_{n,i} \,  \I_{\Nt{i}}
\ee
of $\v_i$ are known at the $i$th terminal, for $i \in \{1,2\}$.

%%%%%%%%%%%%%%%%%%%%%%%%%%%%%%%%%%%%%%%%%%
\section{The proposed closed-form design}
\label{sec:design}

With reference to model \eqref{eq:basic-model},
the
problem at hand is to find optimal values of
$\{\P_i\}_{i=1}^{2}$,
$\F$, and $\{\D_i\}_{i=1}^2$ for recovering
$\s_1$ and $\s_2$ according to a certain
cost function and subject to suitable power constraints
at the terminals and relays.

A common performance measure of the accuracy
in recovering the symbol vector $\s_i$ at terminal $\inot$ is the
mean-square value of the
error $\e_i \eqdef \hat{\s}_i-\s_i$:
$\text{MSE}_i \eqdef \Es[ \| \e_i  \|^2] = \trace(\Reei)$,
where $\Reei \eqdef \Es[\e_i \e_i^\herm ]$
is the error covariance matrix,
which depends on $(\P_{i}, \F, \D_{\inot})$.
As a global cost function for  the overall
two-way transmission,  we consider as in \cite{Lee2010,Ron2011,Hu2013,Hu2016a}
the \emph{sum-MSE}, defined as
$\text{MSE}(\{\P_i\}_{i=1}^{2},\F,\{\D_i\}_{i=1}^{2}) =
\text{MSE}_1  + \text{MSE}_2$.
It is well-known that, for fixed values of
$\{\P_i\}_{i=1}^{2}$ and $\F$,
the matrices $\{\D_i\}_{i=1}^{2}$
minimizing the sum-MSE
are the Wiener filters
\be
\Dmmsei =\C_{i,\inot}^{\herm}
(\C_{i,\inot}\, \C_{i,\inot}^{\herm} + \Rvvi)^{-1}
\label{eq:wiener-flts}
\ee
for $i \in \{1,2\}$, thus yielding
\barr
\text{MSE}(\{\P_i\}_{i=1}^{2},\F) & \eqdef  
\text{MSE}(\{\P_i\}_{i=1}^{2},\F,\{\Dmmsei\}_{i=1}^{2})
\nonumber \\ & =
\sum_{i=1}^{2}
\trace[ ( \I_{\Ns{i}} +
\C_{\inot,i}^{\herm} \Rvvinot^{-1} \C_{\inot,i} )^{-1} ] \: .
\label{eq:MSE-1}
\earr
It is noteworthy that the variables $\P_1$, $\P_2$, and $\F$ are coupled in
\eqref{eq:MSE-1} and, hence, the two terms in \eqref{eq:MSE-1}
cannot be minimized independently.
Herein, we relax the original problem
so as to \textit{separate} the minimization of the two terms in
\eqref{eq:MSE-1}.

As a first step, we observe that minimizing \eqref{eq:MSE-1}
is complicated by the presence of $\Rvvinot^{-1}$,
which depends non-trivially on $\F$.
For such a reason, we consider instead minimization of
the following high signal-to-noise ratio (SNR) approximation:
\be
\text{MSE}(\{\P_i\}_{i=1}^{2},\F) \approx 
\sum_{i=1}^{2}
\trace[ ( \I_{\Ns{i}} +
\sigma^{-2}_{n,\inot} \,
\C_{\inot,i}^{\herm} \C_{\inot,i} )^{-1} ] 
\label{eq:MSE-2}
\ee
which turns out to be accurate when 
$\sigma_w^2 \ll \min(\sigma^2_{n,\inot},\mu_\text{min})$, where
$\mu_\text{min}$ is the smallest eigenvalue of 
$\G_{\inot} \, \F \, \F^\herm \, \G_{\inot}^\herm$.
Suitable constraints must be set
to avoid trivial solutions
in minimizing \eqref{eq:MSE-2}.
It is  customary to impose power
constraints
to limit the average transmit power at the terminals:
\be
\Es[ \| \P_i \s_{i} \|^2] =
\trace(\P_i \, \P_i^\herm) \le \Ptx{i} > 0
\label{eq:TX-const}
\ee
for $i \in \{1,2\}$.
In order to limit $\F$,
we impose a constraint on the average power
received at the terminals in the second time slot,
i.e., with reference to \eqref{eq:rec-1},
we attempt to limit, for $i \in \{1,2\}$,
the following quantities:
\be
\E[ \| \G_i \F \, \y \|^2] =
\trace( \G_i \F \Ryy \F^{\herm} \G_i^{\herm})
\label{eq:RX-const-1}
\ee
where $\Ryy \eqdef \Es[ \y \y^{\herm}] = \sum_{i=1}^{2} \H_i \P_i \P_i^{\herm} \H_i^{\herm} + \sigma_w^2 \I_{\Nc \Nr}$ is the covariance matrix of $\y$.
It is noteworthy that \eqref{eq:RX-const-1} is typically
limited in those scenarios where a target performance has to
be achieved and per-node fairness is not of concern \cite{Beh2008,Cui2004}.
Moreover,  the average power received at the terminals is an important 
metric measuring the human exposure to 
radio frequency  (RF) fields generated by transmitters 
operating at mmWave frequencies \cite{Wu2015}
and, with respect to traditional per-relay transmit power constraints, 
it is more easily related to regulatory specifications \cite{Gastpar2007}. 
To simplify \eqref{eq:RX-const-1}, 
we exploit the following chain of inequalities:
\barr
& \trace( \G_i \F \Ryy \F^{\herm} \G_i^{\herm})
\le
\trace( \G_i \F \F^{\herm} \G_i^{\herm}) \trace(\Ryy) \nonumber \\
& \le
\trace( \G_i \F \F^{\herm} \G_i^{\herm})
\left[ \sum_{i=1}^{2} \trace(\H_i \H_i^{\herm}) \Ptx{i} + \sigma_w^2 \Nc \Nr \right]
\nonumber \\
& \lesssim
\trace( \G_i \F \F^{\herm} \G_i^{\herm})
\Nc \Nr  \left( \sum_{i=1}^{2} \Nt{i} \, \Ptx{i} + \sigma_w^2 \right)
\earr
where the last approximate inequality holds
noting that, for fixed values of $\Nt{i}$,
by the law of large numbers one has
$\H_i^{\herm} \, \H_i /(\Nc \Nr)
\to \I_{\Nt{i}}$
almost surely as $\Nc\Nr$ gets large.
Therefore, if we impose
$\trace( \G_i \F \F^{\herm} \G_i^{\herm}) \le \Prxtilde{i} >0$,
we get the upper bound:
\be
\trace( \G_i \F \Ryy \F^{\herm} \G_i^{\herm}) \lesssim
\underbrace{\Prxtilde{i} \Nc \Nr  \left( \sum_{i=1}^{2} \Nt{i} \Ptx{i} + \sigma_w^2 \right)
}_{\eqdef \Prx{i}} \:.
\ee
Such a choice
allows one to considerably simplify the system design.
In summary, the optimization problem
to be solved can be expressed as
\barr
\min_{\{\P_i\}_{i=1}^{2}, \F} &
\sum_{i=1}^{2}
\trace[( \I_{\Ns{i}} + \sigma^{-2}_{n,\inot} \,
\C_{\inot,i}^{\herm}\C_{\inot,i} )^{-1} ]
\nonumber \\[0.0em]
&
\text{s.to}
\quad
\begin{cases}
\trace(\P_i \, \P_i^\herm) \le \Ptx{i} \\
\trace( \G_{\inot} \F \F^{\herm} \G_{\inot}^{\herm}) \le \Prxtilde{\inot}
\end{cases}
i \in \{1,2\} \: .
\label{eq:prob-1}
\earr
In order to find a closed-form solution of \eqref{eq:prob-1},
we introduce the matrix $\B_i \eqdef \G_i \F \in \setC^{\Nt{i} \times \Nc\Nr}$,
with $i \in \{1,2\}$, and
rewrite \eqref{eq:prob-1} as follows
\barr
\min_{\{\P_i\}_{i=1}^{2}, \{\B_i\}_{i=1}^{2}} &
\sum_{i=1}^{2}
\trace[( \I_{\Ns{i}} + \sigma^{-2}_{n,\inot} \,
\P_{i}^{\herm} \H_{i}^{\herm} \B_{\inot}^{\herm}
\B_{\inot} \H_{i} \P_{i})^{-1} ]
\nonumber \\[0.0em]
&
\text{s.to}
\quad
\begin{cases}
\trace(\P_i \, \P_i^\herm) \le \Ptx{i} \\
\trace( \B_{\inot} \B_{\inot}^{\herm} ) \le \Prxtilde{\inot} \\
\end{cases}
i \in\{ 1,2 \}\: .
\label{eq:prob-2}
\earr
Remarkably, the cost function is the sum of two terms:
the former one depends only on the variables
$\{\P_1,\B_2\}$, whereas the latter one involves only
the variables $\{\P_2,\B_1\}$. Therefore, \eqref{eq:prob-2}
can be decomposed in two problems involving
$\{\P_1,\B_2\}$ and $\{\P_2,\B_1\}$ separately, which can be
solved in parallel in a closed-form manner.
Indeed, capitalizing on such a decomposition,
the solution of \eqref{eq:prob-2}
can be characterized by the following theorem.

\medskip

\begin{Theorem}
\label{th:1}
Assume that:
(a4) $\P_i \in \setC^{\Nt{i} \times \Ns{i}}$ is full-column rank, i.e.,
$\rank(\P_i) = \Ns{i} \leq \Nt{i}$, $i \in \{1,2\}$;
(a5) $\B_{\inot} \H_{i} \in \setC^{\Nt{\inot} \times \Nt{i}}$
is full-column rank, i.e., $\rank(\B_{\inot} \H_{i}) = \Nt{i} \leq \Nt{\inot}$,
for $i \in \{1,2\}$.%
\footnote{(a5) implies
that $\H_i$ is full-column rank too,
i.e., $\rank(\H_i)=\Nt{i}$.}
Moreover, let
$\H_i = \U_{\text{h},i} \bm{\Lambda}_{\text{h},i} \V_{\text{h},i}^{\herm}$
denote the singular value decomposition (SVD) of $\H_i$,
where $\U_{\text{h},i} \in \setC^{\Nc\Nr \times \Nc\Nr}$
and $\V_{\text{h},i} \in \setC^{\Nt{i} \times \Nt{i}}$
are the unitary matrices of left/right singular vectors,
and $\bm{\Lambda}_{\text{h},i} \in \setC^{\Nc\Nr \times \Nt{i}}$
is the rectangular diagonal matrix of the corresponding
singular values arranged in increasing order.
Then, the solution of \eqref{eq:prob-2}
has the following general form:
\barr
\P_{i} &= \V_{\text{h},i,\text{right}} \, \bm{\Omega}_{i} \label{eq:P_sol} \\
\B_{\inot} &= \Q_{\inot} \, \bm{\Delta}_{\inot} \U_{\text{h},i,\text{right}}^{\herm}
\label{eq:B_sol}
\earr
where $\V_{\text{h},i,\text{right}}$ contains the $\Ns{i}$
rightmost columns of $\V_{\text{h},i}$,
$\U_{\text{h},i,\text{right}}$ contains the $\Nt{i}$
rightmost columns of $\U_{\text{h},i}$,
the diagonal matrices
$\bm{\Omega}_i\in \setR^{\Ns{i} \times \Ns{i}}$
and $\bm{\Delta}_{\inot} \in \setR^{\Nt{i} \times \Nt{i}}$
will be specified soon after,
and $\Q_{\inot} \in \setC^{\Nt{\inot} \times \Nt{i}}$
is an arbitrary semi-unitary matrix, i.e.,
$\Q^{\herm}_{\inot} \Q_{\inot} = \I_{\Nt{i}}$.
\end{Theorem}

\begin{proof}
See Appendix \ref{app:proof}.
\end{proof}

\begin{remarkenv}
(a4) implies that
$\Ns{i} \le \Nt{i}$, $i \in \{1,2\}$.
\end{remarkenv}

\begin{remarkenv}
(a5) implies that $\Nt{1} = \Nt{2}$ and, hence,
in the following we set $\Ntone \eqdef \Nt{1} = \Nt{2}$.
\end{remarkenv}

Under (a4) and (a5), the dual-hop channel matrices
$\{\C_{i,\inot} = \B_i \H_{\inot} \P_{\inot}\}_{i=1}^2$
are full-column rank, i.e., $\rank(\C_{i,\inot}) =
\Ns{\inot} \le \Nt{\inot}$, for $i=1,2$: this ensures
perfect recovery of the source symbol vectors
$\{\s_i\}_{i=1}^2$ at the terminals
in the absence of noise
by means of linear equalizers.
Although Theorem \ref{th:1} holds for any value of
$\Ns{1}$ and $\Ns{2}$, we will assume herein
that $\Ns{1}=\Ns{2}=\Ntone$,
which allows the terminals to transmit as many
symbols as possible with an acceptable performance
in practice.

Theorem \ref{th:1} allows one to rewrite
the optimization problem \eqref{eq:prob-2} in a simpler scalar form:
\barr
&\min_{
\substack{
\{ z_{1,\ell}, w_{2,\ell} \}_{\ell=1}^{\Ntone}\\
\{ z_{2,\ell}, w_{1,\ell} \}_{\ell=1}^{\Ntone}
}
}
\sum_{i=1}^{2} \sum_{\ell=1}^{\Ntone}
\frac{1}{1 + \sigma^{-2}_{n,\inot} \, \lambda_{\ell}^2(\H_{i})
z_{i,\ell} \, w_{\inot,\ell}}
\nonumber \\[0.3em]
&
\text{s.to}
\quad
\begin{cases}
\displaystyle \sum_{\ell = 1}^{\Ntone} z_{i,\ell} \le \Ptx{i} \\
\displaystyle \sum_{\ell = 1}^{\Ntone} w_{\inot,\ell} \le \Prxtilde{\inot} \\
w_{\inot,\ell},z_{i,\ell} > 0 & \forall \ell \in \{1,2,\ldots,\Ns{i}\}
\end{cases}
i \in \{1,2\}
\label{eq:prob-3}
\earr
with $z_{i,\ell}$ and $w_{\inot,\ell}$ representing the
$\ell$th squared
diagonal entry of $\bm{\Omega}_{i}$ and $\bm{\Delta}_{\inot}$,
respectively, whereas
$\lambda_{\ell}(\H_{i})$ denotes the $\ell$th nonzero
singular value of
$\H_{i}$, for $\ell \in \{1,2, \ldots, \Ntone\}$.
Similarly to \eqref{eq:prob-2}, problem
\eqref{eq:prob-3} can be decomposed into two separate
problems involving disjoint subsets of variables.

It can be shown, with straightforward manipulations, that
the objective function in \eqref{eq:prob-3} is convex
if and only if
\be
z_{i,\ell} \, w_{\inot,\ell} \geq
\frac{\sigma_{n,\inot}^2}{3\lambda_{\ell}^2(\H_i)}
\label{eq:convex_cond}
\ee
$\forall \ell \in \{1,2,\ldots,\Ns{i}\}$, with $i \in \{1,2\}$.
It is also seen that,
based on (a2),
one has $\lambda_{\text{min}}(\H_i) \gg 1$ in the large $\Nc\Nr$ limit,
with $i \in \{1,2\}$.
Thus, condition \eqref{eq:convex_cond} boils down to
$z_{i,\ell}, \, w_{\inot,\ell} > 0$, for all
$\ell \in \{1,2,\ldots,\Ns{i}\}$, with $i \in \{1,2\}$, which is
already included in the constraints of \eqref{eq:prob-3}.
Therefore, convex programming can be used to find 
a global minimum of \eqref{eq:prob-3}.

To calculate the relaying matrices, let us partition solution
\eqref{eq:B_sol} as $\B_{\inot} = [\B_{\inot,1},\B_{\inot,2},
\cdots,\B_{\inot,\Nc}]$, with
$\B_{\inot,k} \in \setC^{\Ntone \times \Nr}$, $i \in \{1,2\}$.
Defining $\widetilde{\G}_k \eqdef
[\G_{1,k}^\trasp,\G_{2,k}^\trasp]^\trasp
\in \setC^{2 \Ntone \times \Nr}$
and $\widetilde{\B}_k \eqdef
[\B_{1,k}^\trasp,\B_{2,k}^\trasp]^\trasp
\in \setC^{2 \Ntone \times \Nr}$,
and assuming that $\widetilde{\G}_k$ is full-row rank,
i.e., $\rank(\widetilde{\G}_k) = 2 \Ntone \leq \Nr$,
with $k \in \{1,2,\ldots,\Nc\}$, the $k$th relay
can construct its own relaying matrix by solving the matrix equation
$\widetilde{\G}_k \F_k = \widetilde{\B}_k$, whose
minimum-norm solution is given by
\be
\F_k =\widetilde{\G}_k^\dagger  \widetilde{\B}_k
\label{eq:F_sol}
\ee
where the superscript $\dagger$ denotes the Moore-Penrose inverse.

\begin{algorithm}[t]
\caption{The proposed design algorithm}
\small

Input quantities: $\{\H_i,\,\G_i,\,\sigma^2_{n,i},\Ptx{i},\,\Prxtilde{i}\}_{i=1}^2$

Output quantities: $\{\P_i,\Dmmsei\,\}_{i=1}^2$ and $\{\F_k\}_{k=1}^{\Nc}$

\smallskip

\begin{enumerate}

\itemsep=1mm

\item Choose arbitrary $\{\Q_i\}_{i=1}^2$ such that $\Q_i^\herm\Q_i=\I_{\Ns{\inot}}$.

\item Perform the SVD of $\{\H_i\}_{i=1}^2$.
and collect the $\{\Ns{i}\}_{i=1}^2$ largest
singular values and the corresponding left/right singular vectors.

\item Solve the convex problem
\eqref{eq:prob-3}
in the disjoint subsets of variables $\{z_{1,\ell},\,w_{2,\ell}\}_{
\ell=1}^{\Ns{1}}$ and $\{z_{2,\ell},\,w_{1,\ell}\}_{
\ell=1}^{\Ns{2}}$ separately.

\item From the solution of step 3,
build the matrices $\{ \bm{\Omega}_i, \bm{\Delta}_{i} \}_{i=1}^{2}$.

\item Build the matrices $\{\P_i,\,\B_i\}_{i=1}^2$
according to \eqref{eq:P_sol} and \eqref{eq:B_sol}.

\item Calculate $\{\F_k\}_{k=1}^{\Nc}$ according to \eqref{eq:F_sol}.

\item Calculate $\{\Dmmsei\}_{i=1}^2$ according to \eqref{eq:wiener-flts}.
\end{enumerate}

\label{alg:design}
\end{algorithm}

With reference to the step-by-step description of the proposed
design algorithm reported at the top of this page, the following 
comments are in order.
The convex optimization in step 3) can be efficiently carried out
using standard techniques, such as
the interior-point method. 
We observe that the worst-case
theoretical complexity of the
interior-point method  is proportional to
$\sqrt{\Ntone}$.
Hence, for a realistic setting of the system parameters,
the computational complexity of the proposed algorithm,
is dominated by the SVD computation (in step $2$),
which is of order $\mathcal{O}(\Nc \, \Nr \, \Ntone^2)$
and, thus, it \textit{linearly} grows with the number
$\Nc$ of relays.
It is noteworthy that, even though  
the alternating algorithm proposed in \cite{Ron2011}
allows to solve a nonconvex problem by 
solving convex subproblems, 
it is more complex than calculating
the solution of \eqref{eq:prob-3}; moreover, it requires
proper initialization to monotonically converge to  (at least) 
a local optimum.

%%%%%%%%%%%%%%%%%%%%%%%%%%%%%%%%%%%%%%%%%%

\begin{figure}[t]
\begin{center}
\includegraphics[width=\columnwidth]{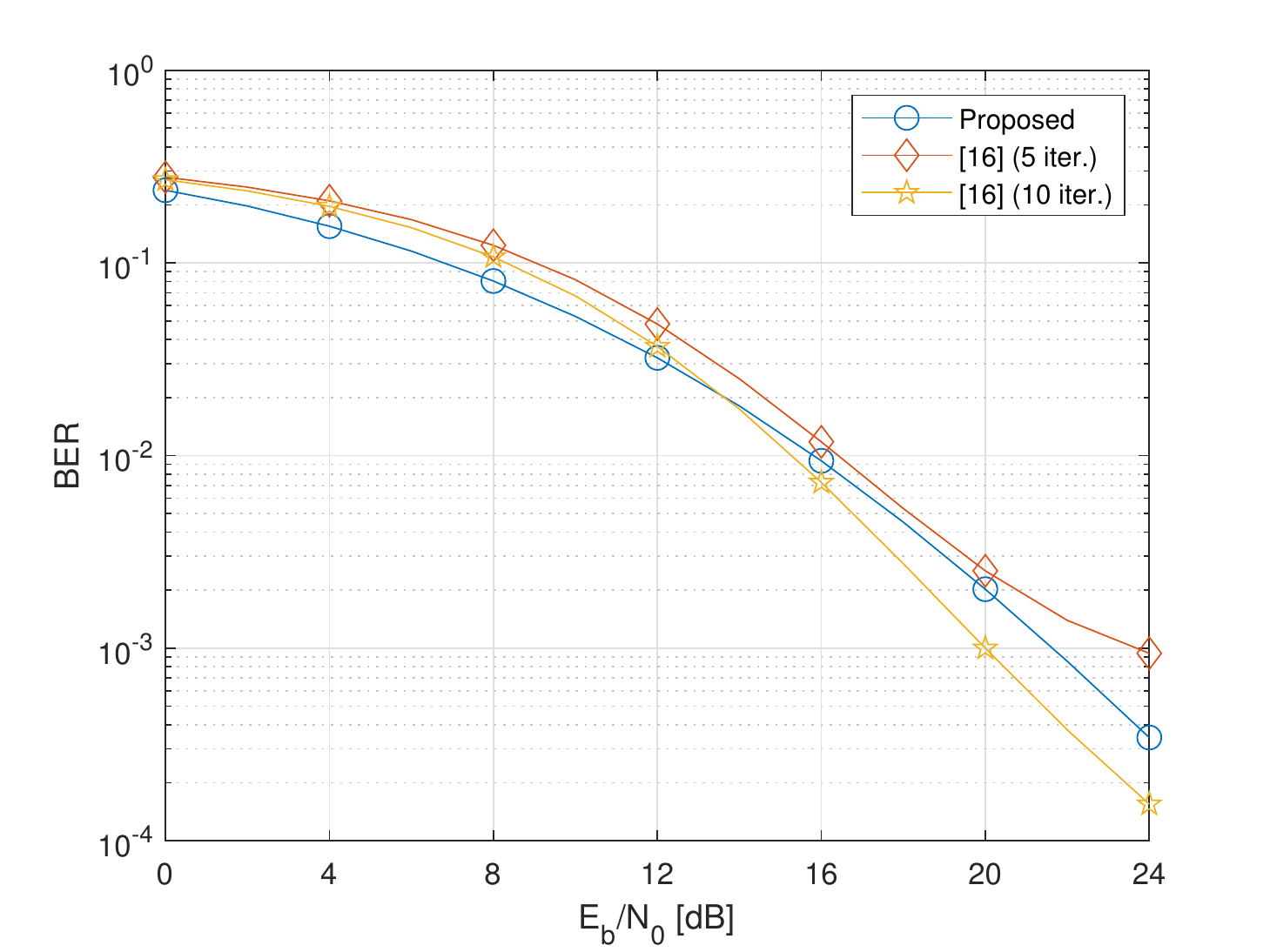}
\caption{BER versus $E_b/N_0$  of the proposed design versus 
the iterative  method of \cite{Ron2011} ($\Nc=2$).}
\label{fig:fig_2}
\end{center}
\end{figure}

\section{Simulation results}
\label{sec:simul}

In this section, to assess the performance of the considered
design, we present the results of Monte Carlo computer simulations,
aimed at evaluating the average (with respect to channel realizations) 
bit-error-rate (BER) of the proposed cooperative two-way MIMO system.
We consider a network encompassing two terminals
equipped with $N_\text{T} = 2$ antennas, and transmitting QPSK symbols with
$\Ns{1}=\Ns{2}=2$.
The $\Nc$ relays are equipped with $\Nr = 4$ antennas.
We also assume that $\Ptx{1}=\Ptx{2}=\Pk=\mathcal{P}$,
for all $k\in\{1,2,\ldots,\Nc\}$, where $\Pk$ represents
the average transmitted power at the $k$th relay,
and set $\sigma^2_{w}=\sigma^2_{n,1}=\sigma^2_{n,2}=1$.
Consequently, the energy per bit to noise power spectral density ratio $E_b/N_0$
is a measure of the per-antenna link quality of both
the first- and second-hop transmissions.
The BER is evaluated by carrying out $10^3$ independent
Monte Carlo trials, with each run using independent sets
of channel realizations and noise, and an independent record
of $10^6$ source symbols.

\begin{figure}[t]
\begin{center}
\includegraphics[width=\columnwidth]{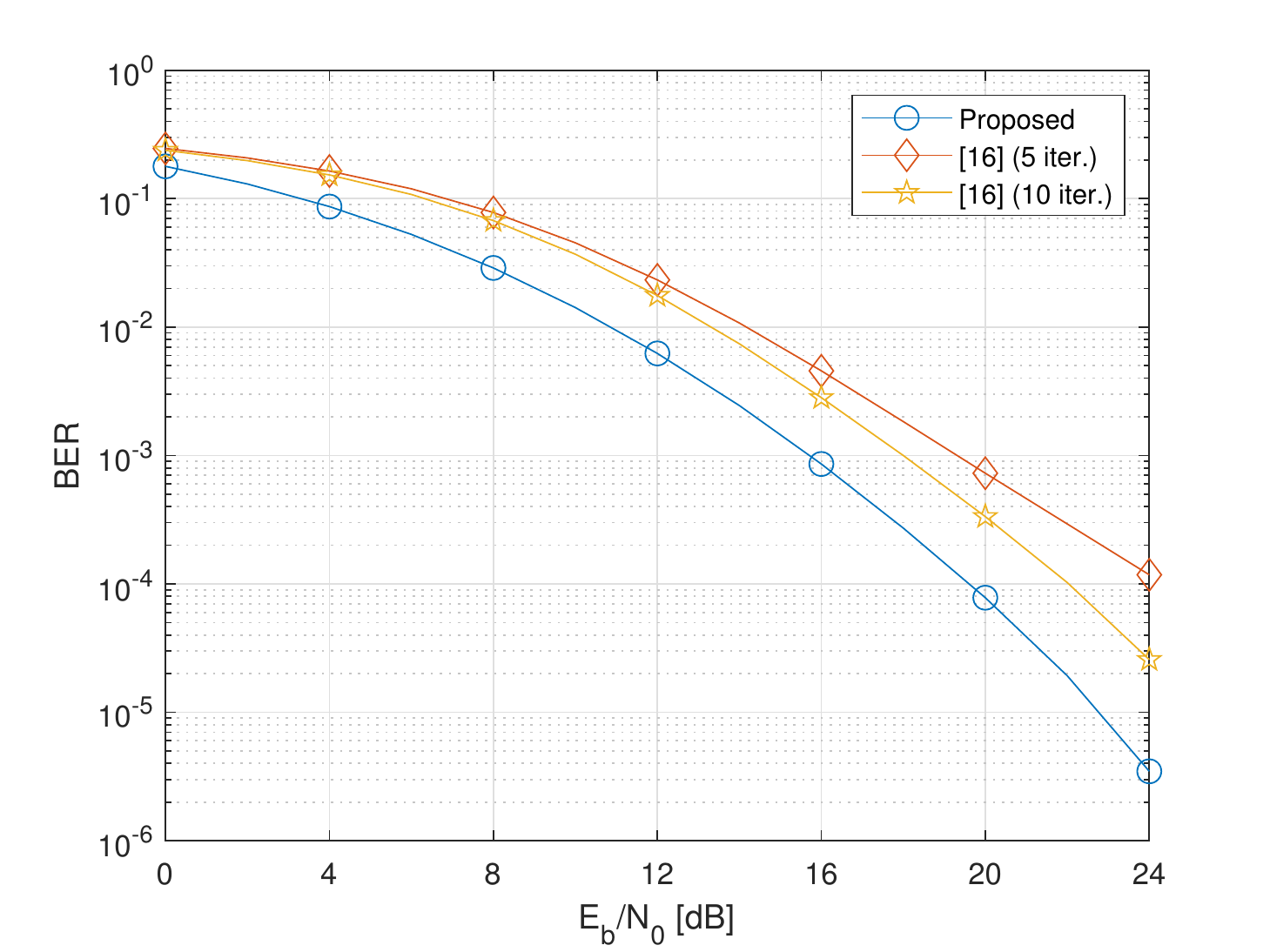}
\caption{BER versus $E_b/N_0$  of the proposed design 
versus the iterative method of \cite{Ron2011} ($\Nc=3$).}
\label{fig:fig_3}
\end{center}
\end{figure}

\begin{figure}[t]
\begin{center}
\includegraphics[width=\columnwidth]{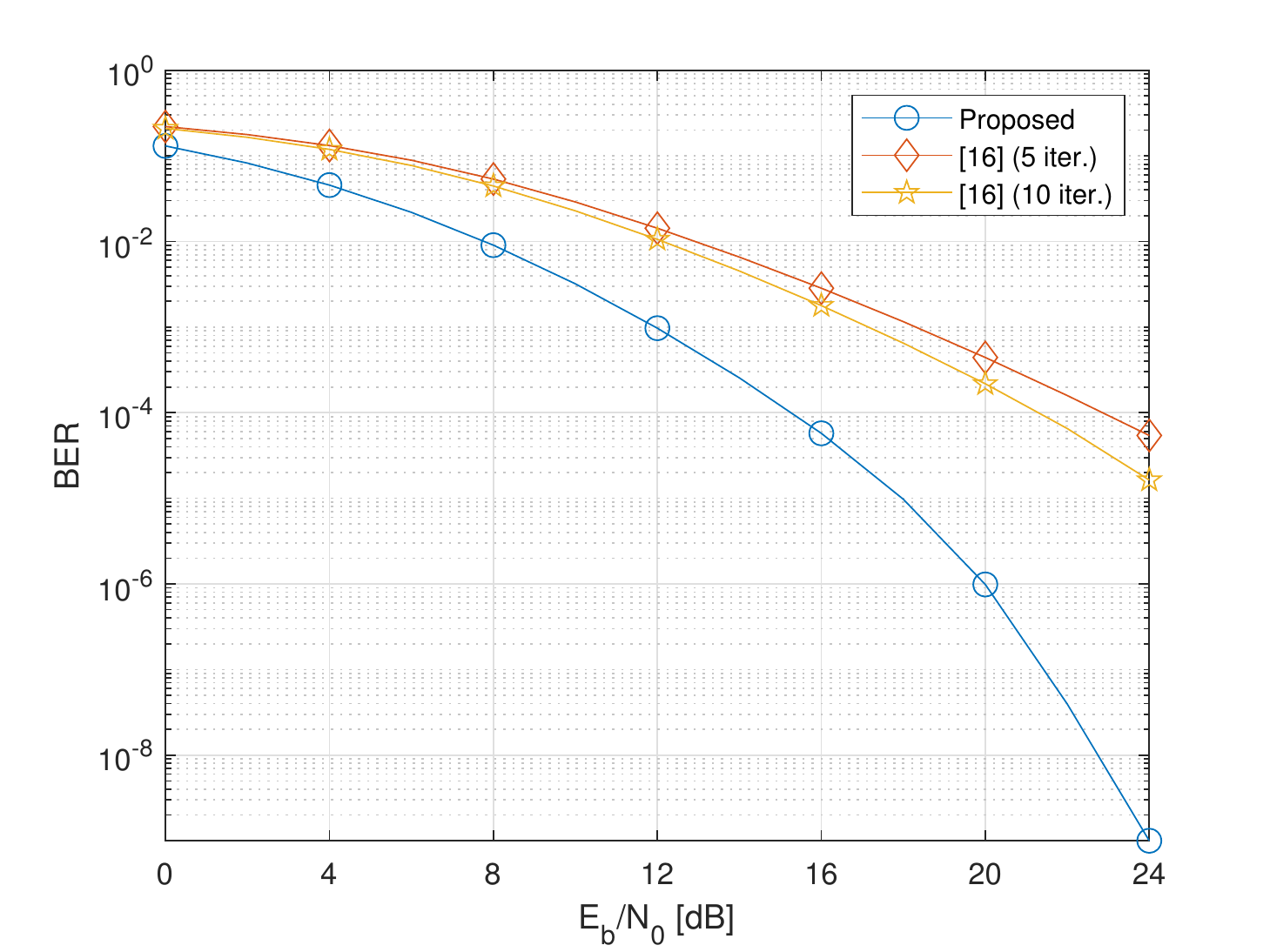}
\caption{BER versus $E_b/N_0$  of the proposed design 
versus the iterative method of \cite{Ron2011} ($\Nc=4$).}
\label{fig:fig_4}
\end{center}
\end{figure}

We compare the performances
of our design (labeled as ``Proposed'')
to those of the
iterative technique proposed in \cite{Ron2011},
which has been shown \cite{Ron2011} in its turn
to outperform other
iterative techniques, such as
the gradient-descent technique
of \cite{Lee2010}.
It is worthwhile to note that both the strategies
under comparison require the same amount of CSI.
Furthermore, since the method of \cite{Ron2011}
imposes different power constraints
on the design of the relaying matrices,
our solutions for $\{\F_k\}_{k=1}^{\Nc}$
are properly scaled
so as to ensure that the average
power transmitted by each relay
is the same for both methods.

In Figs.~\ref{fig:fig_2}--\ref{fig:fig_4},
we report the BER for different values
of the number $\Nc \in \{2, 3, 4\}$ of relays.
Results in Fig.~\ref{fig:fig_2} for $\Nc = 2$
show that the proposed closed-form design,
based on the solution of the relaxed problem
\eqref{eq:prob-2},
exhibits performances comparable with the iterative solution
of \cite{Ron2011} in the considered range of $E_b/N_0$ values
only when the latter employs more than $5$ iterations.
Specifically, when the method of \cite{Ron2011} employs $10$ iterations,
a crossover can be observed in Fig.~\ref{fig:fig_2} between the 
BER curve of the proposed algorithm and that of \cite{Ron2011}. 
This behavior is due to the fact that the rate of convergence of \cite{Ron2011} 
strongly depends on the SNR.
Figs.~\ref{fig:fig_3} and \ref{fig:fig_4} show
that, as the number of relays increases, the proposed method
clearly outperforms the method of \cite{Ron2011} even when the
latter employs $10$ iterations.\footnote{Performance improvement of \cite{Ron2011} 
is negligible after $10$ iterations.}

In a nutshell, although the alternating iterative procedure \cite{Ron2011} 
attempts to solve the nonconvex original two-way constrained minimum sum-MSE
problem, its convergence behaviors are affected in
practice by both the operative SNR and number of relays: in the low-SNR region
and/or when the number of relays is sufficiently large,  convergence to 
a local minimizer is not guaranteed in a reasonable number of
iterations for  all possible initializations.
This is the price to pay for swapping a difficult  joint  optimization  
with a  sequence  of  easier problems involving subsets of the variables.
On the other hand, the proposed optimization strategy 
gives up the idea of solving the original nonconvex problem, 
by resorting to suitable relaxations of both the cost function
and the relaying power constraint. This allows us to 
jointly optimize all the variables, without 
using  burdensome iterative algorithms.

%%%%%%%%%%%%%%%%%%%%%%%%%%%%%%%%%%%%%%%%%%

\section{Discussion and directions for future work}
We tackled the joint sum-MSE design of
terminal precoders/decoders and
relay forwarding matrices for two-way AF MIMO 5G systems.
We showed that a relaxed version of such a problem can be separated into
two simpler ones, which can be solved in parallel by admitting closed-form solutions.
The proposed technique exhibits
a performance gain over the iterative method of \cite{Ron2011},
exhibiting a better scaling with the number of relays and a reduced computational complexity.

In this paper, we assumed the availability of 
full-CSI at both terminals and the relays. 
In this respect, an interesting research subject consists of 
considering the use of partial CSI to extend network lifetime
and reduce the complexity burden. Moreover, since channel
estimation errors occur in practical situations, 
an additional research issue is to develop
robust optimization designs. 

%%%%%%%%%%%%%%%%%%%%%%%%%%%%%%%%%%%%%%%%%%%
\appendixtitles{yes}
\appendix
\section{Proof of Theorem~\ref{th:1}}
\label{app:proof}

We focus on the optimization \eqref{eq:prob-2}
with indexes $i = 1$ and $\inot = 2$, i.e., we consider  
\barr
\min_{\P_1, \B_2} \quad &
\trace[( \I_{\Ns{1}} + \sigma^{-2}_{n,2} \,
\P_1^{\herm} \H_1^{\herm} \B_2^{\herm}
\B_2 \H_1 \P_1 )^{-1} ]
\nonumber \\[0.0em]
&
\text{s.to}
\quad
\begin{cases}
\trace(\P_1 \, \P_1^\herm) \le \Ptx{1} \\
\trace( \B_2 \B_2^{\herm} ) \le \Prxtilde{2} \\
\end{cases}.
\label{eq:prob-separeted}
\earr
We note that under (a4) and (a5), one has
$\rank(\B_2\H_1\P_1) = \Ns{1} \le \Nt{1}$.
Let $\Ua \Lambdaa \Ua^\herm$
be the eigenvalue decomposition (EVD) of
$\A \eqdef \H_1^\herm \B_2^\herm \B_2 \H_1 \in
\setC^{\Nt{1} \times \Nt{1}}$,
where the diagonal matrix $\Lambdaa \in
\setR^{\Nt{1} \times \Nt{1}}$
and the unitary matrix $\Ua \in
\setC^{\Nt{1} \times \Nt{1}}$
collect the eigenvalues, arranged in increasing order,
and the eigenvectors of $\A$, respectively.
The objective function in \eqref{eq:prob-separeted}
is a Schur-concave function of the diagonal elements
of $( \I_{\Ns{1}} + \sigma^{-2}_{n,2} \,
\P_1^{\herm} \A \P_1 )^{-1}$.
In this case, it can be shown \cite{Pal2003c} that
there is an optimal $\P_1$ such that $\P_1^\herm\A\P_1$
is diagonal, whose diagonal elements are assumed
to be arranged in increasing order,
and such an optimal matrix, which also minimizes
$\trace(\P_1\P_1^\herm)$, is given by
\be
\P_1 = \bmath{U}_{\text{a,right}} \bm{\Omega}_1,
\label{eq:P_inc}
\ee
where $\bmath{U}_{\text{a,right}}
\in \setC^{\Nt{1} \times \Ns{1}}$ contains the
$\Ns{1} \le \Nt{1}$ rightmost
columns from $\Ua$, and $\bm{\Omega}_1
\in \setC^{\Ns{1} \times \Ns{1}}$
is a diagonal matrix.
Let $\Q_2 \in \setC^{\Nt{2} \times \Nt{1}}$
be an arbitrary semi-unitary matrix, i.e.,
$\Q_2^\herm \Q_2 = \I_{\Nt{1}}$,
it follows from the EVD of the matrix
$\A$ that $\B_2 \H_1 = \Q_2 \Lambdaa^{1/2} \Ua^\herm$.
Noting that $\rank(\H_1) = \Nt{1}$,
by substituting the ordered SVD of
$\H_1 = \U_{\text{h},1} \bm{\Lambda}_{\text{h},1}
\V_{\text{h},1}^{\herm}$
in this equation, after some algebraic manipulations,
one has that the minimum-norm solution \cite{Ben2003}
of the matrix equation
$\B_2 \U_{\text{h},1} \bm{\Lambda}_{\text{h},1} =
\Q_2 \Lambdaa^{1/2} \Ua^\herm \V_{\text{h},1}$ is
\be
\B_2 =
\Q_2 \Lambdaa^{1/2} \widetilde{\U}_{\text{a}}
\bm{\Lambda}_{\text{h,1,right}}^{-1} \U_{\text{h,1,right}}
\label{eq:B-mat}
\ee
where $\U_{\text{h,1,right}}$ collects the $\Nt{1}$
rightmost columns of $\U_{\text{h},1}$, whereas
$\widetilde{\U}_{\text{a}} \eqdef \Ua^\herm \V_{\text{h},1}
\in \setC^{\Nt{1} \times \Nt{1}}$ and the diagonal
a $\bm{\Lambda}_{\text{h,1,right}} \in
\setR^{\Nt{1} \times \Nt{1}}$ gathers
the $\Nt{1}$ nonzero singular values of $\H_1$
in increasing order.
The aim is now to further determine \eqref{eq:B-mat}
by properly choosing $\widetilde{\U}_{\text{a}}$
such that $\trace(\B_2\B_2^\herm) = 
\trace[( \widetilde{\U}_{\text{a}}
\bm{\Lambda}_{\text{h,1,right}}^{-2}
\widetilde{\U}_{\text{a}} ) \Lambdaa ]$
has the smallest value%
\footnote{It is readily seen that
$\trace(\B_2\B_2^\herm)$ is invariant
to the choice of $\Q_2$.}.
By observing that $\widetilde{\U}_{\text{a}}^\herm
\widetilde{\U}_{\text{a}} = \I_{\Nt{1}}$
and using a known trace inequality, one has
\be
\trace[( \widetilde{\U}_{\text{a}}
\bm{\Lambda}_{\text{h,1,right}}^{-2}
\widetilde{\U}_{\text{a}} ) \Lambdaa ] \geq
\sum_{\ell = 1}^{\Nt{1}} \lambda_{\text{h},1,\ell}^{-2}
\lambda_{\text{a},\ell}
\label{eq:trace-ineq}
\ee
where $\lambda_{\text{h},1,\ell}$ and
$\lambda_{\text{a},\ell}$ denote the
$\ell$th diagonal entry of
$\bm{\Lambda}_{\text{h,1,right}}$
and $\Lambdaa$, respectively.
The equality in \eqref{eq:trace-ineq} holds when
\be
\widetilde{\U}_{\text{a}} =
\Ua^\herm \V_{\text{h},1} = \I_{\Nt{1}}.
\label{eq:trace-eq}
\ee
Substituting \eqref{eq:trace-eq} in \eqref{eq:B-mat},
after some algebraic manipulations,
one obtains $\B_2 = \Q_2 \,
\bm{\Delta}_2 \U_{\text{h,1,right}}^{\herm}$,
with $\bm{\Delta}_2 \eqdef
\Lambdaa^{1/2} \bm{\Lambda}_{\text{h,1,right}}^{-1}
\in \setR^{\Nt{1} \times \Nt{1}}$.
Solution \eqref{eq:P_sol} comes from substituting
in \eqref{eq:P_inc} the minimum-norm solution \cite{Ben2003}
of \eqref{eq:trace-eq}, i.e.,
$\Ua = \V_{\text{h,1,right}}$.

%%%%%%%%%%%%%%%%%%%%%%%%%%%%%%%%%%%%%%%%%%
\vspace{6pt} 

%%%%%%%%%%%%%%%%%%%%%%%%%%%%%%%%%%%%%%%%%%
%% optional
%\supplementary{The following are available online at \linksupplementary{s1}, Figure S1: title, Table S1: title, Video S1: title.}


%%%%%%%%%%%%%%%%%%%%%%%%%%%%%%%%%%%%%%%%%%
\authorcontributions{Conceptualization, D.D., G.G., I.I. and F.V.;
methodology, D.D., G.G., I.I. and F.V.;
writing--original draft preparation, G.G. and F.V.;
writing--review and editing, I.I. and F.V.;
supervision, F.V..}

%%%%%%%%%%%%%%%%%%%%%%%%%%%%%%%%%%%%%%%%%%
%\funding{Please add: ``This research received no external funding'' or ``This research was funded by NAME OF FUNDER grant number XXX.'' and  and ``The APC was funded by XXX''. Check carefully that the details given are accurate and use the standard spelling of funding agency names at \url{https://search.crossref.org/funding}, any errors may affect your future funding.}

%%%%%%%%%%%%%%%%%%%%%%%%%%%%%%%%%%%%%%%%%%
%\acknowledgments{In this section you can acknowledge any support given which is not covered by the author contribution or funding sections. This may include administrative and technical support, or donations in kind (e.g., materials used for experiments).}

%%%%%%%%%%%%%%%%%%%%%%%%%%%%%%%%%%%%%%%%%%
\conflictsofinterest{The authors declare no conflict of interest. The funders had no role in the design of the study; in the collection, analyses, or interpretation of data; in the writing of the manuscript, or in the decision to publish the results.}

%%%%%%%%%%%%%%%%%%%%%%%%%%%%%%%%%%%%%%%%%%
\reftitle{References}


\externalbibliography{yes}
\bibliography{Biblio_two_way_v3}

%%%%%%%%%%%%%%%%%%%%%%%%%%%%%%%%%%%%%%%%%%
\end{document}